\documentclass[aps, reprint, prb]{revtex4-1}
\usepackage{graphicx}
\usepackage[caption=false]{subfig} 
\usepackage{natbib}
\usepackage{multirow}
\usepackage{lipsum}
\usepackage{dcolumn}
\usepackage{amssymb,amsmath}
\usepackage{hyperref}
\hypersetup{colorlinks=true,citecolor=black,urlcolor=blue,linkcolor=black}
\usepackage{enumerate}

\begin{document}
\title{Phase Filters for a Novel Kind of Asymmetric Transport \newline – Scientific Prospects and Opportunities for Possible Applications –}
\author{J. Mannhart, P. Bredol, and D. Braak}
\affiliation{Max Planck Institute for Solid State Research, 70569 Stuttgart, Germany}


\begin{abstract}
We present the concept of nonreciprocal interferometers. These two-way devices let particles pass in both directions, but in one direction break the phase of the particles' wave functions. Such filters can be realized by using, for example, asymmetric quantum rings. Furthermore, we propose arrangements of these interferometers to obtain larger interferometers which are expected to exhibit a puzzling behavior that resembles Maxwell demon action. We indicate an opportunity to resolve this puzzle experimentally.
\par
\vspace{3mm}
\noindent \textit{Submitted 2018/5/5 to Physica E}
\par
\vspace{3mm}
\par
\vspace{3mm}
\noindent \textit{Dedicated to Dr.~Tapash Chakraborty in recognition of his many and valuable contributions to nanoscale physics.}
\end{abstract}

\maketitle

\section{Introduction}

\par Nonreciprocal devices, $i.e.$, devices that let waves pass differently in one direction than in the other, are widely used in radar technology and optics [1]. They have also been implemented, for example, as devices based on plasmons, magnons, electromagnons, and sound waves (see, e.g., [2–6]). Nonreciprocal devices for de-Broglie waves can be realized [7] by using Rashba quantum rings [8,9] or asymmetric Aharonov–Bohm rings [10]. Based on interference of the particles' wave functions, these devices transmit particles preferentially in one direction. We propose nonreciprocal devices that break the symmetry of the phase conservation of objects, depending on the direction in which they pass through the devices. The objects may be particles such as electrons or waves such as photons. Conserving the average energy and momentum of the object, these devices may act in forward direction like open windows, but in reverse direction they may resemble black-body radiators with incoherent output.

\begin{figure}[ht]
	\centering
	\includegraphics[width=0.5 \textwidth]{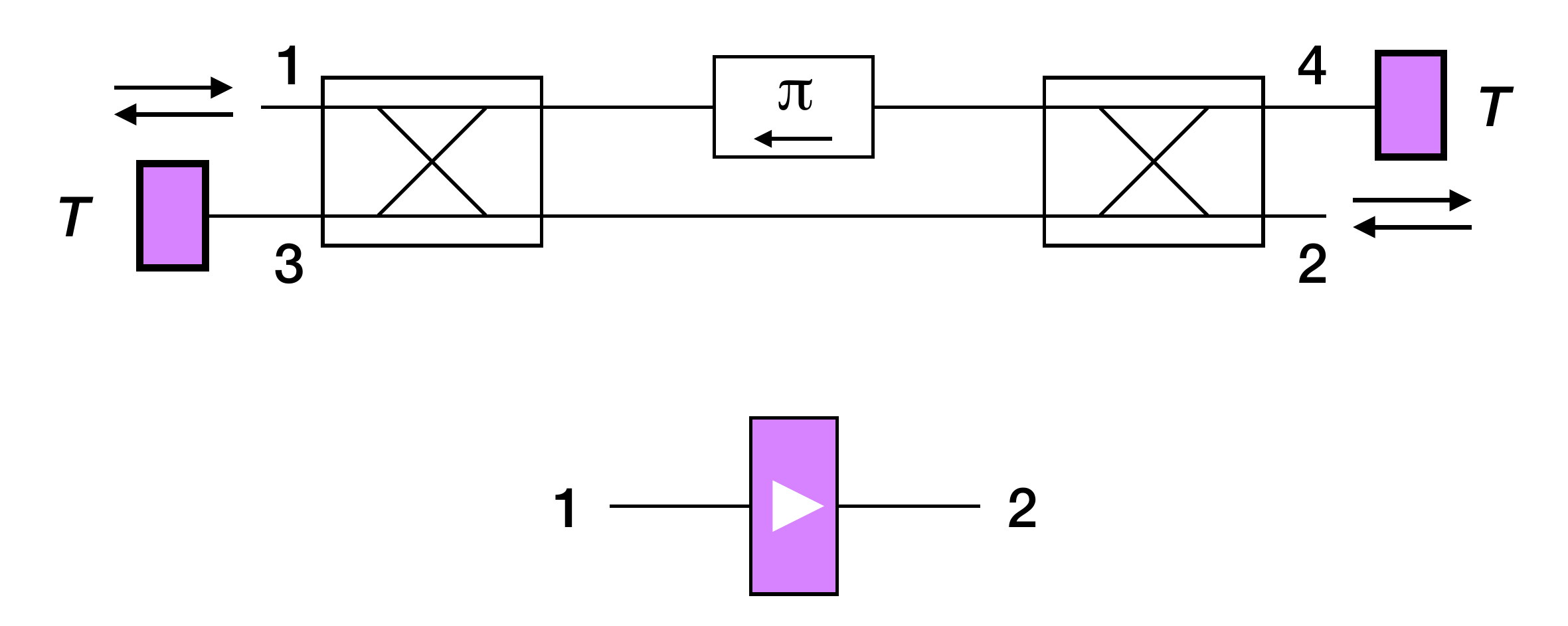}
\caption{Illustration of a four-port circulator in which ports 3 and 4 are terminated by block bodies held at a fixed temperature $T$. The sketch at the bottom is a symbolic representation of the circulators used in Figs. 2 and 3.}
\end{figure}

\section{Implementation of nonreciprocal devices with asymmetric particle coherency}

\par The proposed devices have gyrators [1] as their central component. A gyrator is a nonreciprocal system that lets waves pass in both directions, but adds a phase shift of $\pi$ to the wave function if the wave passes the device in forward direction. A four-port circulator is obtained by combining a gyrator and two hybrid couplers (Fig.~1) [1]. Such a circulator couples an object arriving at the input port $n$ in a cyclic manner to the output port $n+1$, and is thereby characterized by a unitary scattering matrix. We consider the case that ports 3 and 4 are terminated by black-body radiators at a given temperature. An object arriving at port 1 is transmitted without distortion to port 2. If an object is fed into the device at port 2, however, it will be absorbed by the black bodies at ports 3 and 4.  The absorption causes the wave function to collapse, because the dissipative coupling of the absorbed object to the macroscopic bath corresponds to a measurement process. The collapse erases the phase information of the wave. The black bodies absorb all incoming objects and, averaged over timescales characteristic for thermal fluctuations at the given temperature, emit the same amount of energy. The emitted objects, although having on average the same energy as the absorbed objects, carry no memory of the incoming objects' phases. The outgoing objects leave the device at port 1, consistent with the second law of thermodynamics. The phase of an object traveling from 1 to 2 is therefore conserved within its unitary evolution, but the phase of an object leaving the device at port 1 is not correlated with the phase of an object entering at port 2. Moreover, the energy balance of the transfer process is characterized by nonreciprocal behavior. The energy of an object moving on the path from port 1 to 2 is strictly preserved, whereas the energy of particles on the reverse path from 2 to 1 is preserved only on average.
Such devices may be suitable for applications in analog and digital systems because, on statistical average, the devices are transparent to energy and momentum in both directions, but may act as unidirectional membranes for information encoded in the phase of an object. In one direction, such devices transport information without losses; in the other direction they erase information without energy cost (for a valuable review of the problem of information erasure see [11]). 

\section{Asymmetric phase transmission utilized in interference devices}

\par We now consider the possibility of using the asymmetric phase transmission to create asymmetric interference. For this purpose, we arrange two such devices to form a double-slit interferometer (Fig. 2). An object that impinges on such an interferometer at port 1 (left side) passes both slits, thus generating a double-slit interference pattern on the right side (Fig. 2b). In contrast, an object traveling in the port 2 $\rightarrow$ port 1 direction creates two incoherent waves on the left side. These arise from the two independent black-body radiators of the two slits, which radiate according to Lambert's emission law (Fig. 2b). Such an asymmetric behavior results from combining unitary quantum mechanical wave-function evolution, wave-function collapses, and spontaneous emission within the device. The device principles therefore extend beyond quantum mechanics proper. The resulting properties may again be useful for device applications. We present the example shown in Fig. 2c, which consists of the two-slit arrangement incorporated into a cavity, the walls of which are in thermal equilibrium with an external bath.

\begin{figure}[ht]

	\includegraphics[width=0.45 \textwidth]{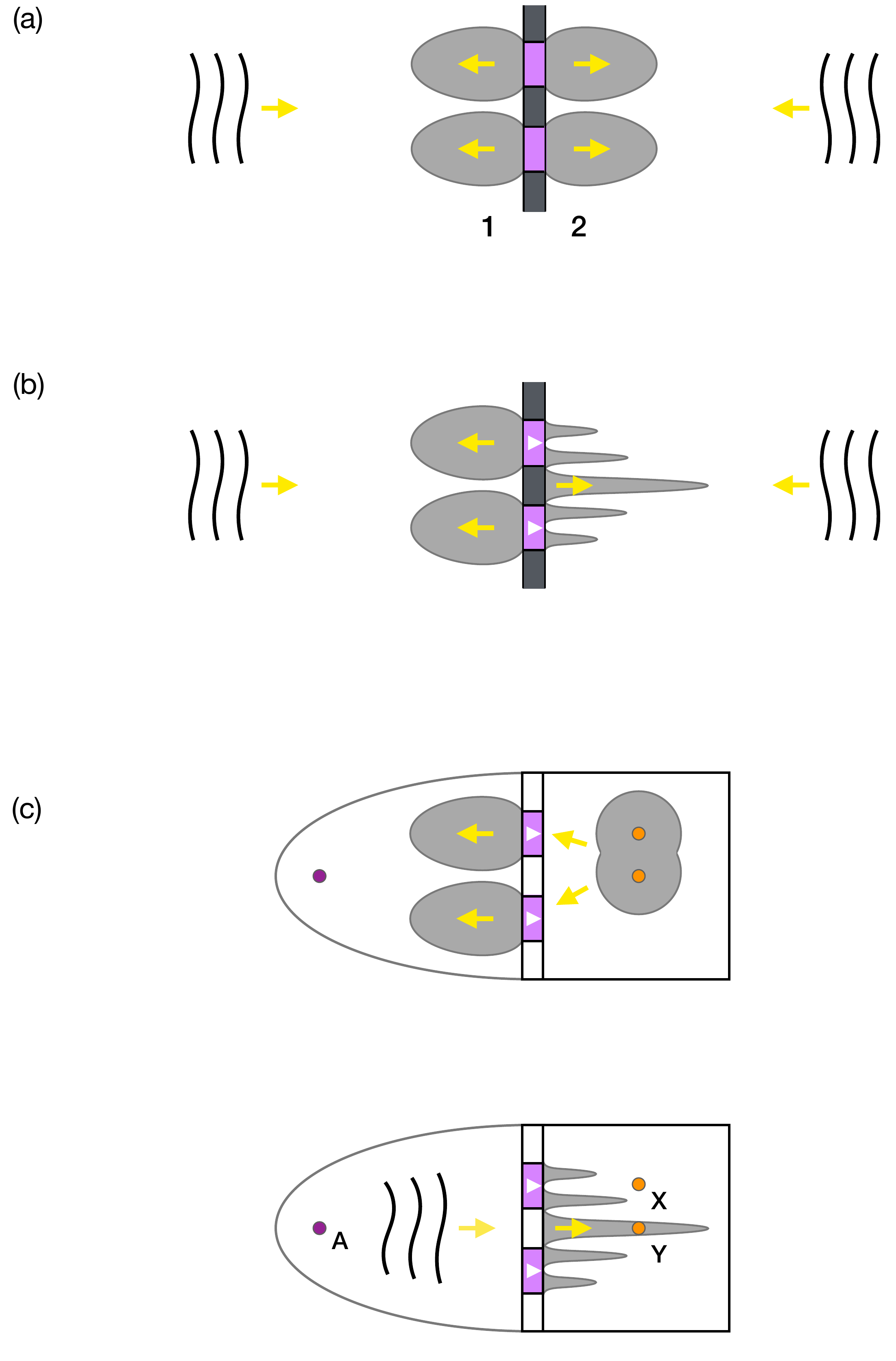}
	\caption{(a) Sketch of coherent radiation (yellow arrows) impinging from two sides on two black-body radiators (pink) arranged in a double-slit configuration. The two black bodies radiate incoherently according to Lambert's law and as symbolized by the gray radiation plumes and yellow arrows. (b) Configuration of (a) in which the double slit is formed by two nonreciprocal interferometers. The double-slit interferometer generates incoherent radiation on the left, but a double-slit interference pattern on the right. (c) Double-slit interferometer of (b) embedded in a cavity. The walls of the left half of the cavity are provided by mirrors; the walls of the right half are black. Some of the radiation coming from the left is generated by spontaneous emission of a small particle that acts as a black-body radiator (particle A, violet). The spatial density of the radiation differs in the two halves of the cavity. The interference characteristic in the right half is inhomogeneous, potentially causing temperature differences of the orange test bodies X and Y placed at different locations in the cavity.}
	\label{F1}
\end{figure}

\par The radiation in the left side of the cavity, which is in thermal equilibrium with the radiation source A, emits a wave from A that is coherent over some angular range, i.e., is phase-correlated. Upon passing the phase filters, this wave therefore generates a diffraction pattern in the right half of the cavity. On the other hand, radiation from the test bodies X and Y will not lead to any such pattern in the left side of the cavity because the phase memory of these waves is erased when the waves pass the phase filters. The radiation now appears to be distributed inhomogeneously in both halves of the cavity in a wavelength-dependent manner. This intensity distribution corresponds to non-maximal entropy and is inconsistent with Kirchhoff's law [12] and the homogeneous distribution assumed in Planck's law of black-body radiation [13]. The generation of the inhomogeneous intensity distribution therefore matches the action of a Maxwell demon [14] and is expected to be inconsistent with the second law of thermodynamics, reminiscent of the devices proposed in [7,15,16]. (Note that the devices described in [15,16] nevertheless agree with the second law.) This puzzle is also apparent if one considers the temperatures of the absorbing bodies X and Y, each of which is in internal thermal equilibrium and positioned inside and outside the diffraction peaks in the right half cavity. Measurements of the diffraction patterns, for example by photodetectors, will provide experimental access to whether a wave-function collapse is a purely epistemic concept or a physical phenomenon, which, in the setup presented here, would contradict the second law.

\par If, as the arguments suggest, experiments or a full theoretical treatment revealed that the proposed devices break the second law, the implications for science would be significant, independent of the devices' hypothetical practical impact. For practical applications, it would be advantageous to aim for individual devices on a microscopic length scale and to incoherently combine these devices in large numbers. If, for example, the effect were generated by asymmetric, ring-shaped organic molecules, with each molecule producing statistically only a minuscule fraction of the thermal energy k$T$ per electron passing through the ring, a mole of these molecules could possibly achieve a macroscopic output.

\section{Summary and conclusion}

\par In this short contribution dedicated to Dr. Tapash Chakraborty, we have proposed nonreciprocal devices that show an asymmetric phase transport of waves, including matter waves. The phases of waves passing the devices in forward direction are conserved. For waves traversing the devices in the opposite direction, the waves' phases are replaced by random ones. If incorporated into interferometers, this phase-breaking asymmetry causes asymmetric diffraction, thus yielding devices that may constitute Maxwell demons, thereby being inconsistent with the second law of thermodynamics. In principle, these devices may be realizable, possibly even at room temperature and above, and therefore have scientific and technological implications. Opportunities are wide open for further theoretical and experimental explorations to resolve the open questions. 

\begin{acknowledgments}
We gratefully acknowledge valuable support by L.-M. Pavka and helpful discussions with H. Boschker, T. Kopp, and C. Schön. D.B. thanks Raymond Fresard for the warm hospitality during his stay at CRISMAT, ENSICAEN, and for support by the French Agence Nationale de la Recherche, Grant No. ANR-10-LABX-09-01 and the Centre national de la recherche scientifique  through Labex EMC3.
\end{acknowledgments}

\end{document}